\documentclass[aps,pra,preprint,groupedaddress,showpacs,showkeys]{revtex4}
\usepackage{graphicx}
\begin{document}

\title{Drying model for porous material based on the dynamics of the evaporation front}

\author{Reza Torabi, Mohammad Mehrafarin}
\email{mehrafar@aut.ac.ir}
\affiliation{Physics Department, Amirkabir University of Technology, Tehran 15914,
Iran}

\date{\today}

\begin{abstract}
A receding-front model for drying of porous material is proposed that explains their drying-rate curves based on the dynamics of the evaporation front. The falling-rate regime is attributed to the slowing down of the front's propagation inside the medium due to the resistance offered by the disorder generated by porosity. The model is solved numerically and the resulting drying-rate curve is obtained for the falling-rate period. The curve shows a linear behavior at early times in conformance with experiment. 
\end{abstract}


\maketitle

\section{Introduction}
Drying refers to the process of removal of free moisture from the surface as well as the interior of a solid material by the application of heat. The output from a drying experiment can be plotted as a `drying-rate curve', which is a graph of the drying rate, $Q=-A^{-1} dw/dt$ (moisture loss per unit time, $t$, per unit area, $A$, of the exposed drying surface), versus moisture content, $w$ (moisture mass per unit mass of the dry solid) of the material. For porous material in general, a common feature of such curves is the existence of two well-defined regimes, namely, a constant drying-rate period succeeded by a falling-rate stage that is at first linear \cite{Coulsont,Mujumdar}. The constant-rate period, wherein a film of free moisture is always available at the drying surface, is governed fully by the transport conditions external to the material. At the so-called critical moisture content that marks the end of the first regime, the drying rate begins to fall since the interior moisture, due to internal transport limitations, can no longer migrate at sufficient rate to the drying surface to keep it saturated. The evaporation front, thence, begins to recede from the drying surface into the solid; moisture reaching the surface via migration through the pore volume. Thus, during the falling-rate period, the internal transport mechanisms govern the drying process.

Numerous contributions to the theories of drying rate, with particular emphasis on the falling-rate regime, can be found in the literature (see \cite{ Coulsont,Mujumdar,Waananeu,Mehrafarin} and the references therein, in particular \cite{Waananeu} which provides an extensive bibliography of references dealing with drying models for porous solids). The present work introduces yet another drying model that lies in the category of the so called receding-front models (see also \cite{Palancz,Sadykov}). However, rather than considering the heat and mass transfer as is customarily done, we shall take a different {\it physical} approach based on the dynamics of the evaporation front according to the following picture. The evaporation front defines a moving interface between the wet and dry regions. It is formed by the moisture film surface all through the first stage of drying. Consequently, as long as the external conditions remain unchanged, the velocity of the interface and hence the drying rate will be constant. However, as the front penetrates inside the porous medium during the second stage, the disorder in the medium acts as an inhomogeneous friction force, opposing the overall advance of the interface. The slowing down of the evaporation front propagation, thence, results in the fall of the drying rate. The disorder, generated by the randomness of the pore space and the solid matrix within the porous material, does not change with time and, therefore, can be represented as a quenched noise. Quenched noise has been widely applied to model such disorder and its nontrivial effect on the motion and morphology of an interface is well known \cite{Buldyrev,Csahok1,Csahok2, Amaral,Barabasi}.

After introducing the drying model in the next section, we solve it numerically to obtain the drying-rate curve for the falling-rate stage. The curve shows a linear behavior at early times in conformance with experiment. The linear falling-rate regime is described in terms of the two free parameters of the model that mimic the independent effects of the porosity and the external conditions of the material.

\section{The drying model}
As long as a moisture film is available at the drying surface, the evaporation front will lie at its free interface. Consequently, provided external conditions do not change, the velocity of the interface and hence the drying rate will remain constant. This trivially reconciles with the constant-rate period of the drying-rate curve. Let us, therefore, focus on the falling-rate period wherein the front forms a rough interface as it penetrates through the medium \cite{Shaw}.

\begin{figure}
\caption{\label{fig1}Evaporation front in porous medium}
\includegraphics{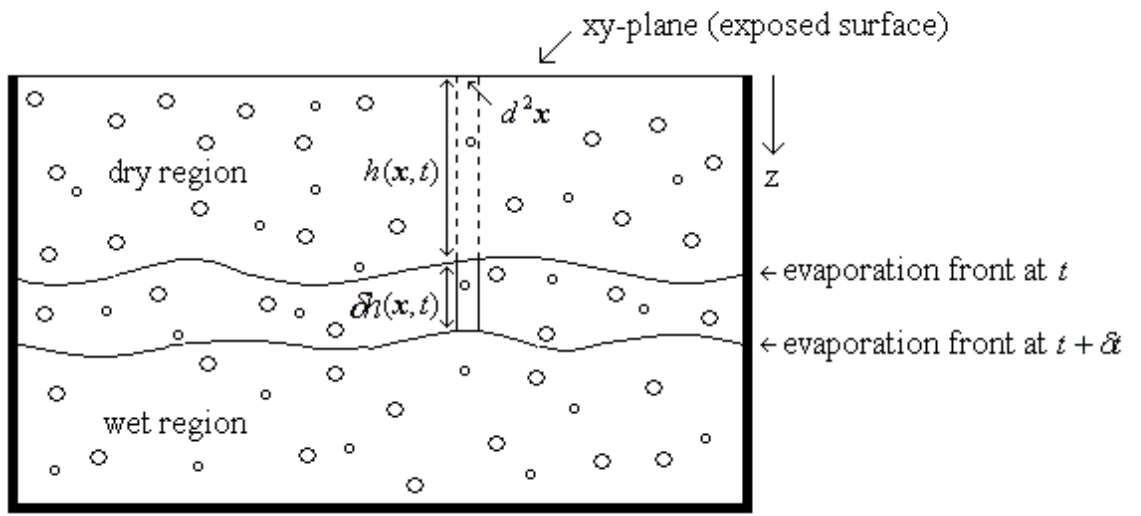}
\end{figure}

We consider a preparation for the drying material, wherein the exposed surface is flat and the cross-sections parallel to this surface are uniform (FIG. \ref{fig1}). Take the plane of the exposed surface as the $xy$-plane and let $z=h({\bf x},t)$ denote the height of the evaporation front below the surface at position ${\bf x}\equiv (x,y)$ and time $t$. Consider an areal element of the front whose projection onto the $xy$-plane is $d^2x\equiv dxdy$. As the front recedes inside the medium, this element sweeps out a volume per unit time given by $(\partial_t{h})d^2x$. This corresponds to a {\it pore} volume per unit time equal to $\varepsilon({\bf x},z=h) \partial_t {h} d^2x$, where, $0\leq\varepsilon({\bf x},z)\leq 1$ represents the local porosity (pore volume per unit total volume) of the medium. Of course, due to random fluctuations, the evolution of the front is stochastic; that is, each repetition of the experiment from the same initial conditions may result in a different final configuration for the interface. Let us, therefore, denote the average over the ensemble of all such configurations by angular brackets. Since moisture resides inside pore volume, the rate of change of the moisture content will, thus, be given by
$$\dot {w}= -\frac{\rho}{M}\int_R <\varepsilon({\bf x},h) \partial_t {h}> d^2x,$$
where, $\rho$ is the density of moisture (water), $M$ is the mass of the dry solid and the integral is over the whole exposed region, $R$. Hence, we have for the drying rate, that,
\begin{equation}
Q=\frac{\rho}{MA}\int_R <\varepsilon({\bf x},h) \partial_t {h}> d^2x. \label{one}
\end{equation}
As a first approximation in treating the disorder, which is generated by the (random) spatial variation of the porosity, we shall take the inhomogeneity into account only in so far as it affects the dynamics of the front (i.e., $\partial_t {h}$). That is, we write (\ref{one}) as 
\begin{equation}
Q=\frac{\rho \varepsilon}{M}\ v, \label{two}
\end{equation}
where, $\varepsilon$ is now the mean porosity and
$$v(t)=\frac{1}{A}\ \int_R <\partial_t {h}> d^2x, $$
is the average velocity of the interface. Equation (\ref{two}) is the central equation of our model, which relates the drying rate to the propagation velocity of the evaporation front. 

To find the propagation velocity, we need to model the time variation of the interface height at any position ${\bf x}$. The propagation of the front is induced by the flux of moisture particles escaping the interface from random positions. This random flux causes the height to increase locally at a rate that fluctuates around some average value $V$. Because of the uncorrelated nature of the randomness in the escape process, these fluctuations are described by the `thermal noise', $\eta_t({\bf x},t)$, which is defined through
\begin{equation}	
<\eta_t({\bf x},t)>=0, \ \ \ \  <\eta_t({\bf x},t) \eta_t({\bf x}^\prime,t^\prime)>=2D_t \delta^2({\bf x}-{\bf x}^\prime) \delta(t-t^\prime). \label{noise}
\end{equation}
On top of this, we must add the effect of the disorder generated by the random inhomogeneity of the porosity of the medium, which directly affects the local velocity of the interface by acting as an inhomogeneous friction force. Hence, the dynamics can be described by the equation
\begin{equation}	
\frac{\partial {h({\bf x},t)}}{\partial {t}} = V+ \eta_t({\bf x},t)+ \eta_q({\bf x},z=h), \label{three} 
\end{equation}
where, the quenched noise, $\eta_q({\bf x},z)$, represents the disorder as mentioned in the introduction and is defined through
\begin{equation}	
<\eta_q({\bf x},z)>_0=0, \ \ \ \  <\eta_q({\bf x},z) \eta_q({\bf x}^\prime,z^\prime)>_0=2D_q \delta^2({\bf x}-{\bf x}^\prime) \delta(z-z^\prime). \label{four}
\end{equation}
Of course, in (\ref{four}), the averages (not to be confused with $<...>$ and hence the notation) are over the different realizations of the randomness; for a given material, the quenched disorder is to be represented through fixing a particular realization. Whence, as far as the quenched randomness is concerned, the evolution of the interface is deterministic, i.e., restarting from the same initial conditions always results in the same final configuration. The stochastic element enters the dynamics solely through the thermal noise.

Equation (\ref{three}) is by far the simplest equation describing the motion of the evaporation front in a porous medium and together with (\ref{two}), therefore, defines our drying model for the (linear) falling-rate regime. These two equations yield the following for the drop in the drying rate:
\begin{equation}	
\Delta Q(t)\equiv Q(0)-Q(t)= \frac{\rho \varepsilon}{M}\ \frac{1}{A} \int_R \{ \eta_q({\bf x},0) -<\eta_q({\bf x},h({\bf x},t))> \} d^2x, \label{five}
\end{equation}
where, $t=0$ represents the start of the falling-rate regime (corresponding to a flat interface at the drying surface, i.e., $h({\bf x},0)=0$). Also, because $<\partial_t{h}>= \partial_t{<h>}$ from equation (6), (\ref{two}) can be integrated at once to obtain the moisture loss according to
\begin{equation}	
\Delta w(t) \equiv w_c - w(t)=\frac{\rho \varepsilon}{M}AH, \label{six}
\end{equation}
where, $w_c$ is the critical moisture content and
$$H(t)=\frac{1}{A}\int_R <h({\bf x},t)> d^2x, $$
is the average height of the interface. Thus, at the start of the falling-rate regime $w$ has the initial value $w_c$, while decreasing as the time elapses at a rate which is directly determined by the velocity of the evaporation front. Once the configuration of the interface at arbitrary time, $t$, is determined from its equation of motion, (\ref{five}) and (\ref{six}) can be used to obtain the drying-rate curve as a graph of $\Delta Q$ versus $\Delta w$. 

\section{Numerical solution and results}
We introduce the dimensionless variables
$$\tilde{\bf{x}}\equiv \frac{1}{\ell_0}{\bf x},\ \ \tilde{z}\equiv \frac{z}{\ell_0},\ \ \tilde{t}\equiv \frac{t}{t_0},$$
where, the characteristic length and time scales are defined by the two coefficients entering the equation of motion according to the combinations 
$$\ell_0^3 \equiv \frac{D_t^2}{D_q},\ \ t_0^3\equiv \frac{D_t^5}{D_q^4}.$$
These two independent scales are set by the disorder in the medium (through $D_q$) as well as the external conditions that influence the escape process (through $D_t$); most notably the temperature of the hot environment. They remain free parameters of the model to be determined ultimately from comparison of the results with experiment. For numerical purposes, it is useful to introduce the dimensionless quantities
$$\tilde{h}(\tilde{{\bf x}},\tilde{t})\equiv \frac{1}{\ell_0} h({\bf x},t),\ \ \tilde{V}\equiv \frac{V}{V_0},\ \ \tilde{\eta}_t(\tilde{{\bf x}},\tilde{t})\equiv \frac{1}{V_0} \eta_t({\bf x},t),\ \ \tilde{\eta}_q(\tilde{{\bf x}},\tilde{z})\equiv \frac{1}{V_0} \eta_q({\bf x},z),$$
where,
$$ V_0\equiv \frac{\ell_0}{t_0}=\frac{D_q}{D_t},$$
is the characteristic velocity. Using (\ref{noise}) and (\ref{four}), the second moments of the noise variables, therefore, translate into
\begin{equation}	
<\tilde{\eta}_t(\tilde{{\bf x}},\tilde{t}) \tilde{\eta}_t(\tilde{{\bf x}}^\prime,\tilde{t}^\prime)>=2\ \delta^2(\tilde{{\bf x}}-\tilde{{\bf x}}^\prime) \delta(\tilde{t}-\tilde{t}^\prime),\ \ <\tilde{\eta}_q(\tilde{{\bf x}},\tilde{z}) \tilde{\eta}_q(\tilde{{\bf x}}^\prime,\tilde{z}^\prime)>_0=2\ \delta^2(\tilde{{\bf x}}-\tilde{{\bf x}}^\prime) \delta(\tilde{z}-\tilde{z}^\prime),
\end{equation}
while the first moments remain zero, of course. The stochastic equation of motion, thus, becomes
\begin{equation}
\frac{\partial {\tilde{h}(\tilde{{\bf x}},\tilde{t}})}{\partial {\tilde{t}}} = \tilde{V}+ \tilde{\eta}_t(\tilde{{\bf x}},\tilde{t})+ \tilde{\eta}_q(\tilde{{\bf x}},\tilde{h}), \label{seven}
\end{equation}
with $\tilde{h}(\tilde{{\bf x}},0)=0$, as the initial condition. To integrate this equation numerically, we discretize it by taking the points $(\tilde{{\bf x}}, \tilde{z})$ to belong to a cubic lattice of size $L^3$ and use the Euler algorithm with time steps $\Delta \tilde{t}$. (The lattice points should be much more closely spaced in the $\tilde{z}$ direction than in the other two directions. This is because the $\tilde{z}$ coordinate of the quenched noise in (\ref{seven}) is pre-determined by the value of $\tilde{h}$.) The noises $\tilde{\eta}_t(\tilde{{\bf x}},\tilde{t})$ and $\tilde{\eta}_q(\tilde{{\bf x}},\tilde{z})$ are both chosen from a uniform distribution of random numbers between $-1$ and $+1$; a distribution that, clearly, satisfies the above conditions on the moments. However, the quenched noise is generated only once and for all (corresponding to fixing a particular realization of the disorder, as pointed out before), while the thermal noise is regenerated at every repetition of the calculation. Hence, we generally end up with a different value for $\tilde{h}(\tilde{{\bf x}},\tilde{t})$ at each repetition.

\begin{figure}
\caption{\label{fig2}Rescaled drying-rate curve in the falling-rate regime}
\includegraphics{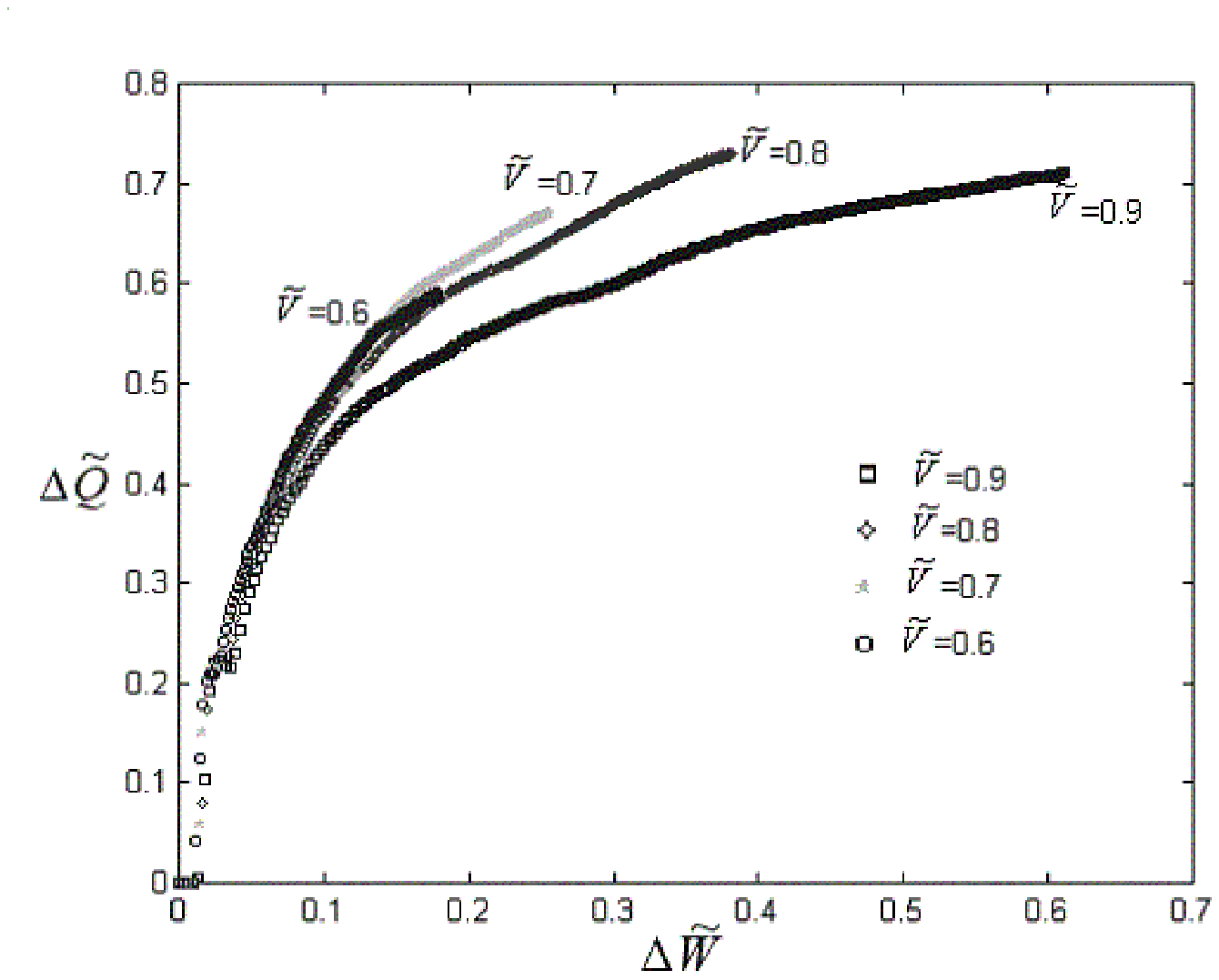}
\end{figure}

In terms of the rescaled (dimensionless) rate drop and the rescaled moisture loss,
\begin{equation}
\Delta \tilde{Q}(\tilde{t})\equiv \frac{M}{\rho \varepsilon V_0}\Delta Q(t),\ \ \ \Delta \tilde {w}(\tilde{t}) \equiv \frac{M}{\rho \epsilon \ell_0 A} \Delta w(t), \label{rescale}
\end{equation}
respectively, equations (\ref{five}) and (\ref{six}), thus, reduce to
\begin{eqnarray}
\Delta \tilde{Q}(\tilde{t})=L^{-2} \sum_{\tilde{{\bf x}}} \{\tilde{\eta}_q(\tilde{{\bf x}},0)-<\tilde{\eta}_q(\tilde{{\bf x}},\tilde{h}(\tilde{{\bf x}},\tilde{t}))>\} \nonumber\\
\Delta \tilde {w}(\tilde{t})=L^{-2} \sum_{\tilde{{\bf x}}} <\tilde{h}(\tilde{{\bf x}},\tilde{t})>\ \ \ \ \ \ \ \ \ \ \ \ \, 
\end{eqnarray}
In FIG. \ref{fig2}, we plot the rescaled drying-rate curve as a graph of $\Delta \tilde{Q}$ versus $\Delta \tilde {w}$ for various values of $\tilde{V}$ and $L=40$. (Variation of $L$ did not produce noticeable change.) It is seen that the curves all collapse into a single straight line at early times, thereby establishing a `universal' linear behavior. Using (\ref{rescale}) one can, therefore, fit any experimental drying-rate curve during the {\it linear} falling-rate period by appropriate adjustment of the independent parameters $\ell_0$ and $t_0$. Our model, thus, describes the linear falling-rate regime in terms of two free parameters that represent the independent effects of the porosity and the external conditions of the material.


\end{document}